\documentclass[aps, prd,
preprint,
superscriptaddress,
preprintnumbers,
nofootinbib,
showpacs,
citeautoscript
]{revtex4}

\IfFileExists{srcltx.sty}{\usepackage[active]{srcltx}}

\def\be{\begin{equation}}%
\def\ee{\end{equation}}%
\def\ba{\begin{eqnarray}}%
\def\ea{\end{eqnarray}}%
\usepackage{graphicx}
\usepackage{amsfonts,amssymb,amsmath} %

\usepackage{xspace} %

\IfFileExists{slashed.sty}{\usepackage{slashed}}

\ifx\slashed\undefined %
\def\slashed#1{\gamma^\mu #1_\mu} %
\else \fi

\newcommand{\eqn}[1]{&\hspace{-0.3em}#1\hspace{-0.3em}&}

\newcommand{\numsm}{{\ensuremath{\nu}MSM}\xspace}

\begin{document}
\title{
  \begin{flushright}
    \small \textnormal{CERN-PH-TH/2006-006}
  \end{flushright}%
  The masses of active neutrinos in the $\nu$MSM from X-ray astronomy}
%
\author{A. Boyarsky}
\affiliation{CERN, Theory department, Ch-1211 Geneve 23,
  Switzerland}
  
\author{A. Neronov} 
\affiliation{ INTEGRAL Science Data Center, Chemin d'\'Ecogia 16,
  1290 Versoix, Switzerland and Geneva Observatory, 51 ch. des Maillettes,
  CH-1290 Sauverny, Switzerland}
  
  \author{O. Ruchayskiy}
\affiliation  {Institut des
  Hautes \'Etudes Scientifiques, Bures-sur-Yvette, F-91440, France}
  
  \author{M. Shaposhnikov}
\affiliation  {\'Ecole Polytechnique F\'ed\'erale de Lausanne, Institute of
  Theoretical Physics, FSB/ITP/LPPC, BSP 720, CH-1015, Lausanne,
  Switzerland}

\date{\today}
%
\begin{abstract}
  In an extention of the Standard Model by three relatively light right-handed
  neutrinos (the $\nu$MSM model) the role of the dark matter particle is
  played by the lightest sterile neutrino. We demonstrate that the
  observations of the extragalactic X-ray background allow to put a strong
  upper bound on the mass of the lightest active neutrino and predict the
  absolute values of the mass of the two heavier active neutrinos in the
  \numsm, provided that the mass of the dark matter sterile neutrino is
  larger than $1.8$ keV.
\end{abstract}

\maketitle

Recently a simple extention of the Minimal Standard Model (MSM) by three
relatively light (with the Majorana masses smaller than the electroweak scale)
right-handed neutrinos was suggested~\cite{Asaka:2005an,Asaka:2005pn}. The
model (dubbed \emph{$\nu$MSM}) allows simultaneous explanation of neutrino
oscillations and baryon asymmetry of the universe and proposes the candidate
for the dark matter -- the lightest long-lived sterile neutrino. Unlike
traditional see-saw mechanism~\cite{Seesaw}, the tiny values of the active
neutrino masses in this model are related to the small Yukawa coupling
constants between sterile neutrinos and left-handed leptonic doublets. Small
Yukawas are essential for the explanation of dark matter: the lightest sterile
neutrino with the mass of a few keV can be sufficiently long-lived
\cite{Dodelson:1993je,Dolgov:2000ew}. In addition to this, small Yukawa
couplings are crucial for baryogenesis in the $\nu$MSM, leading to coherence
in the oscillations of sterile neutrinos \cite{Akhmedov:1998qx,Asaka:2005pn}.
Moreover, they are required by some models, explaining pulsar kick
velocities~\cite{Kusenko:1997sp}.

The Lagrangian of the \numsm has the following form:
\begin{equation}
  \label{eq:2}
  \mathcal{L}_{\numsm} = \mathcal{L}_{MSM}+i\bar N^I \slashed{\partial} N_I -
  \Bigl(\bar L_\alpha M^D_{\alpha I} N_I + \frac12 M_I \bar N_I^c N_I + h.c.\Bigr)\;.
\end{equation}
In Eq.~(\ref{eq:2}) we fixed the basis in which the sterile neutrino Majorana
mass matrix $M_I$ is diagonal (fields $N_I$ denote right-handed neutrinos,
$I=1,2,3$), whereas the Dirac mass matrix providing the mixing between
left-handed $L_\alpha$ and right-handed neutrinos is $M^D_{\alpha I} \equiv
F_{\alpha I}\langle \Phi\rangle$ where $\alpha = \{e,\mu,\tau\}$ is the active
neutrino flavor, $F_{\alpha I}$ is a set of Yukawa couplings, and $\langle
\Phi\rangle \simeq 174$~GeV is the Higgs vacuum expectation value.  The
sterile neutrino with $I=1$ is supposed to be the lightest one.

It was pointed out in \cite{Asaka:2005an} that a prediction of the absolute
values of the masses of active neutrinos can be made, provided that the only
source of the sterile neutrino production in the early universe is their
mixing with active neutrinos.  Assuming that the initial concentration (say,
at temperatures greater than 1~GeV) of sterile neutrinos were zero, one can
estimate~\cite{Dodelson:1993je,Dolgov:2000ew,Abazajian:2001nj} the sterile
neutrino abundance $\Omega_s$ and identify it with that of the dark matter
$\Omega_{DM}$.  The abundance is proportional to the square of the Dirac mass
and depends only weakly on the sterile neutrino Majorana mass in the keV
region \cite{Dodelson:1993je,Dolgov:2000ew,Abazajian:2001nj}. With the use of
these results the constraint on the parameters of the $\nu$MSM can be written
as \cite{Asaka:2005an}: 
\be 
\sum_{\alpha = e,\mu,\tau} |M^D_{\alpha 1}|^2= m_0^2 \,,
\label{DMCond}
\ee 
where $m_0^2 = {\cal O}(0.1 \mbox{eV})^2$ with an uncertainty of,
say, factor of a few, coming from the poor knowledge of the dynamics
of the hadronic plasma at the temperature of the sterile neutrino
production.  The restrictions from the observations of the cosmic
microwave background and the matter power spectrum inferred from
Lyman-$\alpha$ forest data~\cite{Hansen:2001zv,Viel:2005qj} gives a
constraint (see also \cite{Abazajian:2005xn} for a most recent
study): 
\begin{equation} 
  M_1 > 2~ \mbox{keV}\;.
  \label{lyman}
\end{equation}
Similarly to Eq. (\ref{DMCond}) this constraint assumes that the sterile
neutrino was produced in active-sterile neutrino oscillations and plays a role
of so-called warm dark matter (WDM). A weaker, but an assumption-free lower
bound on the mass of the lightest sterile neutrino \be M_1 \gtrsim
0.5~\mbox{keV}
\label{dwarf}
\ee
comes from the application of Tremaine-Gunn arguments
\cite{neutrinoDM} to the dwarf spheroidal galaxies \cite{Lin:1983vq}.
Now, since $M_1$  from (\ref{lyman},\ref{dwarf}) is much larger than
$m_0$, the see-saw formula \cite{Seesaw} for active neutrino masses 
\begin{eqnarray}
  \label{eq:Mseesaw}
  M^\nu = -\left(M^D\right)^T \; M_I^{-1} \; M^D 
\end{eqnarray}
is valid. The analysis of Eq.~(\ref{eq:Mseesaw}) reveals
\cite{Asaka:2005an} that the constraint (\ref{DMCond}) together with
(\ref{lyman}) leads to an upper bound on the lightest neutrino mass,
$m_\nu < m_0^2/M_1 \simeq {\cal O}(10^{-5})$~eV. Now, since this
bound is much smaller than $\sqrt{\Delta m_{\rm sol}^2}\simeq
10^{-2}$ eV, where $\Delta m_{\rm sol}^2$ is the solar mass square
difference, the masses of other two active neutrinos
should be given
by 
\be
m_2 = \sqrt{\Delta m_{\rm sol}^2},~~~
m_3 =\sqrt{\Delta m_{\rm atm}^2}
\label{normal}
 \ee
or by 
\be
m_1 \approx m_2 = \sqrt{\Delta m_{\rm atm}^2},~~
m_1^2-m_2^2 = \Delta m_{\rm sol}^2~,
\label{inverted}
\ee
if the hierarchy is inverted\footnote{The same conclusion is true if
the bound (\ref{dwarf}) is taken.}. In numbers
\cite{Strumia:2005tc}, 
\be
\Delta m_{\rm sol}^2=(7.2-8.9)\cdot 10^{-5}~\mbox{eV}^2,~~~
\Delta m_{\rm atm}^2=(1.7-3.3)\cdot 10^{-3}~\mbox{eV}^2~.
\label{exp}
\ee
The errors correspond to 99\% confidence level
range of $2.58 \sigma$. Clearly, the predictions
(\ref{normal},\ref{inverted}) remain in force provided the mass of
the lightest neutrino is smaller than the error bar in the solar
neutrino mass difference, namely for (99\% C.L.)
\be
m_\nu< 3\cdot 10^{-3} \mbox{eV}~.
\label{condition}
\ee 

In fact, the results (\ref{DMCond}) and (\ref{lyman}) are not
universal and do depend on the details of universe evolution above
the temperature of Big Bang Nucleosynthesis (BBN). In particular,
they are sensible to the universe content at the time of sterile
neutrino production and on particle physics well beyond the
electroweak scale. For example, a substantial lepton asymmetry may
lead to enhancement of the production of the sterile neutrinos
\cite{Shi:1998km}. If the reheating temperature of the universe is
just above the nucleosynthesis scale \cite{Gelmini:2004ah} the
consideration leading to Eq.~(\ref{DMCond}) is not applicable at all.
If the sterile neutrino has large enough coupling to inflaton it will
be created right after inflation rather than at small temperatures.
The reheating of the universe between the moment of the sterile
neutrino production and nucleosynthesis due to late phase transitions
or due to hypothetical heavy particle decays would dilute the
concentration of the sterile neutrinos and decrease their momentum,
leading to smaller free streaming lengths at the onset of structure
formation.

In this note we consider the question whether a {\em robust}
prediction (which does not depend on the uncertainties discussed
above) can be made for active neutrino masses in the $\nu$MSM
provided all 100 \% of the dark matter in the universe is associated
with sterile neutrino. 

Our analysis is based on the astrophysical constraint coming from the
analysis of the X-ray background derived in \cite{astro} (for earlier
works see \cite{Dolgov:2000ew,Abazajian:2001vt}). In the $\nu$MSM the
lightest sterile neutrino can decay into active neutrino and photon
with the width given by a trivial generalization of Pal and
Wolfenstein formula~\cite{Pal:1981rm,sar}:
\begin{eqnarray}
    \label{eq:G_nuga}
 \Gamma_{\gamma}
  \eqn{=} \frac{9\, \alpha_{\rm em} \, G_F^2 \, M_{1}^3}{256 \, \pi^4}
  \sum_{\alpha = e, \mu, \tau} |M^D_{\alpha 1}|^2 \,.
  \end{eqnarray}
The increase of the Dirac neutrino mass $|M^D_{\alpha 1}|^2$ (notice
that exactly the same combination appears in Eq.~(\ref{DMCond}))
would lead to the increase of the X-ray flux from the sterile
neutrino dark matter.  Clearly, the astrophysical constraints on this
flux would lead to the limit on $m_0$, and, therefore to the
prediction of active neutrino masses if $m_0$ happens to be small
enough.  

The corresponding bound on $m_0$ can be found from ref. 
\cite{astro}.  As we have shown in this paper, a non-observation of a
peculiar feature in the X-ray background associated with the decays
of dark matter sterile neutrino implies that\footnote{ Sterile
neutrino mass $M_1$ was denoted by $m_s$ in~\cite{astro}, mixing
angle $\theta$ is defined via $\tan 2\theta \equiv 2 m_0/M_1$.}
\begin{equation}
  \label{eq:1}
  \Omega_s\sin^2(2\theta)<3\times 10^{-5}
  \left(\frac{M_1}{\mathrm{keV}}\right)^{-5}~,
\end{equation}
where $\Omega_s$ is the dark matter abundance.  This equation
describes an empirical fit to the corresponding exclusion region
coming from the analysis of the HEAO-1 and XMM missions and is valid
at $M_1> 1$ keV (see \cite{astro} and references therein). In the
limit $m_0\ll M_1$, one gets an upper bound on the lightest active
neutrino mass $m_\nu < m_0^2/M_1$. Combined with Eq.~(\ref{eq:1}),
this can be recast into the constraint on $m_\nu$:
\begin{equation}
  \label{eq:3}
  m_\nu < 3.4\times 10^{-2}\left(\frac{0.22}{\Omega_s}\right)
  \left(\frac{\mathrm{keV}}{M_1}\right)^{4}~\text{eV}~.
\end{equation}
The weakest limit on the mass of the lightest active neutrino comes
from the region of smaller sterile neutrino masses.  One can see that
for $\Omega_s = 0.22$ and 
\be
M_1 > 1.8~ \mbox{keV}
\label{mcrit}
\ee
the condition (\ref{condition}) is satisfied. In other words, the
prediction of active neutrino masses (\ref{normal},\ref{inverted}),
made in~\cite{Asaka:2005an} is robust if the mass of dark matter
sterile neutrino is large enough. 

In conclusion, we have demonstrated that in the $\nu$MSM the astronomical
observations of the X-ray background allow to put severe constraints on the
mass of the lightest sterile neutrino and to make a prediction of the masses
of other active neutrinos, independently on assumptions on the evolution of
the early universe above the BBN temperatures, provided the mass of dark
matter neutrino is larger than $1.8$ keV. For smaller masses $M_1$, admitted
by the constraint (\ref{dwarf}), the predictions (\ref{normal},\ref{inverted})
are not in general valid.

The authors thank A. Kusenko for discussion.  This work was supported
in part by the Swiss Science Foundation and by European Research
Training Network contract 005104 "ForcesUniverse".


\begin{thebibliography}{99}
\bibitem{Asaka:2005an} T.~Asaka, S.~Blanchet and M.~Shaposhnikov,
  Phys.\ Lett.\ B {\bf 631} (2005) 151
  [arXiv:hep-ph/0503065].

\bibitem{Asaka:2005pn}
T.~Asaka and M.~Shaposhnikov,
Phys.\ Lett.\ B {\bf 620}, 17 (2005)
[arXiv:hep-ph/0505013].

  
\bibitem{Seesaw}
P.~Minkowski,
Phys.\ Lett.\ B {\bf 67} (1977) 421;
T.~Yanagida, in {\em Proc. of the Workshop on the Unified Theory and the
  Baryon Number in the Universe}, Tsukuba, Japan, Feb.~13-14, 1979, p.~95,
eds. O.~Sawada and S.~Sugamoto, (KEK Report KEK-79-18, 1979, Tsukuba); Progr.\ 
Theor.\ Phys.\ {\bf 64} (1980) 1103 ;
M.~Gell-Mann, P.~Ramond and R.~Slansky, 
in {\em Supergravity}, 
eds. P.~van~Niewenhuizen and D.~Z.~Freedman
(North Holland, Amsterdam 1980);
P.~Ramond, 
in {\em Talk given at the Sanibel Symposium}, 
Palm Coast, Fla., Feb.~25-Mar.~2, 1979, preprint CALT-68-709
(retroprinted as hep-ph/9809459);
S.~L.~Glashow,
in {\em Proc. of the Carg\'ese  Summer Institute on Quarks and Leptons},
Carg\'ese, July 9-29, 1979, 
eds. M.~L\'evy et. al, , (Plenum, 1980, New York), p707;
R.~N.~Mohapatra and G.~Senjanovic,
Phys.\ Rev.\ Lett.\  {\bf 44} (1980) 912.

 \bibitem{Dodelson:1993je}
S.~Dodelson and L.~M.~Widrow,
Phys.\ Rev.\ Lett.\  {\bf 72} (1994) 17 
[arXiv:hep-ph/9303287].

\bibitem{Dolgov:2000ew}
A.~D.~Dolgov and S.~H.~Hansen,
Astropart.\ Phys.\  {\bf 16} (2002) 339 
[arXiv:hep-ph/0009083].

\bibitem{Akhmedov:1998qx}
E.~K.~Akhmedov, V.~A.~Rubakov and A.~Y.~Smirnov,
Phys.\ Rev.\ Lett.\  {\bf 81} (1998) 1359 
[arXiv:hep-ph/9803255].

\bibitem{Kusenko:1997sp}
  A.~Kusenko and G.~Segre,
  Phys.\ Lett.\ B {\bf 396} (1997) 197
  [arXiv:hep-ph/9701311].

\bibitem{Abazajian:2001nj}
K.~Abazajian, G.~M.~Fuller and M.~Patel,
Phys.\ Rev.\ D {\bf 64} (2001) 023501 
[arXiv:astro-ph/0101524].


\bibitem{Hansen:2001zv}
  S.~H.~Hansen, J.~Lesgourgues, S.~Pastor and J.~Silk,
  Mon.\ Not.\ Roy.\ Astron.\ Soc.\  {\bf 333} (2002) 544
  [arXiv:astro-ph/0106108].

\bibitem{Viel:2005qj}
M.~Viel, J.~Lesgourgues, M.~G.~Haehnelt, S.~Matarrese and A.~Riotto,
Phys.\ Rev.\ D {\bf 71} (2005) 063534
[arXiv:astro-ph/0501562].


\bibitem{Abazajian:2005xn}
  K.~Abazajian,
  arXiv:astro-ph/0512631.

  
\bibitem{neutrinoDM} S. Tremaine and J. E. Gunn, 
  Phys. Rev. Lett. \textbf{42}, 407 (1979).
  
\bibitem{Lin:1983vq}
  D.~N.~C.~Lin and S.~M.~Faber,
  Astrophys.\ J.\  {\bf 266} (1983) L21.


\bibitem{Strumia:2005tc}
  A.~Strumia and F.~Vissani,
  Nucl.\ Phys.\ B {\bf 726} (2005) 294
  [arXiv:hep-ph/0503246].
  
\bibitem{Shi:1998km}
  X.~d.~Shi and G.~M.~Fuller,
  Phys.\ Rev.\ Lett.\  {\bf 82}, 2832 (1999)
  [arXiv:astro-ph/9810076].

\bibitem{Gelmini:2004ah}
G.~Gelmini, S.~Palomares-Ruiz and S.~Pascoli,
Phys.\ Rev.\ Lett.\  {\bf 93}, 081302 (2004)
[arXiv:astro-ph/0403323].


\bibitem{astro} A. Boyarsky, A. Neronov, O. Ruchayskiy and M. Shaposhnikov,
  astro-ph/0512509.

\bibitem{Abazajian:2001vt}
K.~Abazajian, G.~M.~Fuller and W.~H.~Tucker,
Astrophys.\ J.\  {\bf 562} (2001) 593 
[arXiv:astro-ph/0106002].

\bibitem{Pal:1981rm}
P.~B.~Pal and L.~Wolfenstein,
Phys.\ Rev.\ D {\bf 25}, 766 (1982).

\bibitem{sar} V.~D.~Barger, R.~J.~N.~Phillips and S.~Sarkar,
  Phys.\ Lett.\ B {\bf 352}, 365 (1995)
  [Erratum-ibid.\ B {\bf 356}, 617 (1995)]
  [arXiv:hep-ph/9503295].

\end{thebibliography}
\end{document}